\documentclass[conference,a4paper]{IEEEtran}
\IEEEoverridecommandlockouts
\usepackage{stfloats,color}
\usepackage{amsfonts}
\usepackage[cmex10]{amsmath}
\usepackage{graphicx}
\usepackage{setspace}
\usepackage{cite}
\usepackage{array}
\usepackage{algorithmic}
\usepackage{mdwmath}
\usepackage{mdwtab}
\usepackage{url}
\usepackage[center]{caption}
\usepackage{fixltx2e}
\usepackage{url}
\usepackage{times}
\usepackage{epsfig}
\usepackage{latexsym}
\usepackage{epstopdf}
\usepackage{verbatim}
\usepackage{units}
\usepackage{amsthm}
\usepackage{mdwlist}
\usepackage{balance}
\theoremstyle{remark}

\graphicspath{{./figures/}}

\begin{document}
\title{Power-Optimal Feedback-Based Random Spectrum Access for an Energy Harvesting Cognitive User
\thanks{This work was made possible by NPRP grants 6-1326-2-532 and 4-1034-2-385 from the Qatar National Research Fund (a member of Qatar Foundation). The statements made herein are solely the responsibility of the authors.}
}

\author{\IEEEauthorblockN{Mahmoud Ashour$^\dag$, Ahmed El Shafie$^{*\ddag}$, Amr Mohamed$^\dag$ and Tamer Khattab$^\ddag$}

\IEEEauthorblockA{$^\dag$ Computer Science and Engineering Dept., Qatar University, Doha, Qatar.\\
$^*$ Wireless Intelligent Networks Center (WINC), Nile University, Giza, Egypt.\\
$^\ddag$ Electrical Engineering Dept., Qatar University, Doha, Qatar.\\
}
\{m.ashour, ahmed.elshafie, amrm, tkhattab\}@qu.edu.qa
}

\maketitle

\begin{abstract}
In this paper, we study and analyze cognitive radio networks in which secondary users (SUs) are equipped with Energy Harvesting (EH) capability. We design a random spectrum sensing and access protocol for the SU that exploits the primary link's feedback and requires less average sensing time. Unlike previous works proposed earlier in literature, we do not assume perfect feedback. Instead, we take into account the more practical possibilities of overhearing unreliable feedback signals and accommodate spectrum sensing errors. Moreover, we assume an interference-based channel model where the receivers are equipped with multi-packet reception (MPR) capability. Furthermore, we perform power allocation at the SU with the objective of maximizing the secondary throughput under constraints that maintain certain quality-of-service (QoS) measures for the primary user (PU).

\end{abstract}


\section{Introduction}

Cognitive radio is a promising technology that efficiently utilizes the radio spectrum \cite{zhao2007survey}. Its essence resides in allowing secondary users (SUs) to access the spectrum while the primary users (PUs) are silent. Moreover, secondary nodes can use the spectrum concurrently with primary nodes in condition that they guarantee certain quality-of-service measures for the primary network.

Cognitive radio networks may involve some battery-based nodes and hence, energy management becomes a very important aspect. In many applications, the SU is a battery-constrained device. Therefore, its operation, which includes spectrum sensing and access, depends largely on its energy resources. Recently, energy harvesting has been gaining increasing worldwide interest. It is an emerging technology that is considered a promising solution to wireless energy-constrained networks. The significance of this technology resides in its capability to prolong wireless networks lifetime \cite{c0}. With such technology, nodes are capable of collecting energy from the surrounding environment \cite{c1,survey}.

Studying data transmission by an energy harvester with a rechargeable battery has received a lot of attention in the literature \cite{lei2009generic,sharma2010optimal,ho2010optimal,yang2010transmission,yang2010optimal,tutuncuoglu2010optimum,pappas2011optimal,krikidis2012stability}. Using a dynamic programming framework, Lei \emph{et al.} \cite{lei2009generic} derived the optimal online policy for controlling admissions into the data buffer. In \cite{sharma2010optimal}, Sharma \emph{et al.}
investigate the energy management policies that stabilize the data queue of a single-user communication. The authors derive some delay-optimal properties. For energy harvesting systems in
a time-constrained slotted setting, the authors in \cite{ho2010optimal} discussed the throughput-optimal energy allocation. In \cite{yang2010transmission,yang2010optimal}, Yang \emph{et al.} minimize the transmission completion time of an energy harvesting system. The optimal solution is obtained using a geometric framework. In \cite{tutuncuoglu2010optimum}, the authors solve the problem of maximizing the amount of data transmitted in a finite time horizon. The network under consideration consists of energy harvesting transmitters with batteries of finite energy storage.

In the context of cognitive communications, the authors in \cite{pappas2011optimal} consider a simple network composed of one PU and one SU. The PU is assumed to be equipped with a rechargeable battery (energy queue with random arrivals). The energy queue is assumed to be modelled as a decoupled ${\rm M/D/1}$ queue with Bernoulli arrivals and unity service rate. The SU is plugged to a reliable power supply; it always has energy/power. The maximum stable throughput region was derived using the dominant system approach \cite{dominance_approach}.

The optimal sensing and access policies for an energy harvesting SU based on Markov decision processes (MDPs) have been investigated in \cite{Sultan}. In \cite{ElSh1312:Optimal}, the optimal sensing duration of an energy harvesting SU is selected randomly at the beginning of every time slot from a predetermined set. The stable throughput is obtained via optimizing over the possibility of choosing certain sensing durations.


Due to the broadcast nature of wireless channels, primary feedback can be overheard by all nodes in the network. The main problem with schemes that employ spectrum sensing only is that sensing does not inform the SU about the impact of its transmissions on the primary receiver. This issue induced interest in leveraging the feedback sent by the primary receiver at the end of time slots to the primary transmitter in order to optimize the secondary transmission strategies. In \cite{eswaran2007bits}, Eswaran \emph {et al.} assume that the SU observes the feedback from the primary
receiver as it reflects the achieved primary rate. The SU aims at maximizing its throughput while guaranteeing a certain primary packet rate. The authors of \cite{lapiccirella2010cognitive} use a partially observable Markov decision process (POMDP) to optimize the
secondary action using the spectrum sensing outcome and primary Automatic Repeat reQuest (ARQ) feedback.\cite{huang2010distributed} investigates the secondary power control based on primary feedback. In \cite{wimob} and \cite{ourletter}, El Shafie \emph{et al.} derive the maximum stable throughput of a rechargeable SU sharing the channel with a PU. The SU perfectly overhears primary feedback.

In this paper, we analyze the performance of an energy harvesting SU sharing the spectrum with a PU. We propose a new random spectrum sensing and access scheme based on the outcome of the primary feedback signals overheard by the SU as well as the SU's spectrum sensing quality.
Unlike previous works proposed earlier in literature, we investigate the impact of erroneous primary feedback signals on the performance of both the PU and SU.

Our contributions in this paper can be summarized as follows. We assume feedback errors at the SU while receiving the primary feedback signals. In contrast to most of the literature, e.g., \cite{eswaran2007bits,lapiccirella2010cognitive,huang2010distributed,wimob,ourletter}, we investigate the impact of feedback errors on the secondary throughput as well as the average primary packet delay. Moreover, in contrast to \cite{wimob}, we consider Poisson energy arrivals at the energy queue and perform power allocation at the SU to achieve the maximum throughput. We propose a new spectrum sensing and access based on the joint outcome of primary feedback reception at the SU and the spectrum sensing process. We optimize over the sensing and access probabilities such that the secondary throughput is maximized, subject to the constraint that the primary queue is kept stable. We then study the worst case scenario for the PU and append a queueing delay constraint to the optimization problem that carries further quality-of-service (QoS) guarantees for the PU.

The rest of this paper is organized as follows. Section \ref{system_model} presents the system model along with the proposed spectrum access protocol. In Section \ref{performance_analysis}, we present the queueing analysis of both PU and SU and derive a closed-form expression for the average primary packet delay. Next, we derive lower and upper bounds on the secondary throughput in Section \ref{lower_and_upper_bounds}. Section \ref{numerical_results} shows our numerical results. Finally, concluding remarks are drawn in Section \ref{conclusion}.

\section{System Model}\label{system_model}
We consider the cognitive radio system shown in Fig. \ref{Fig1}.
The system consists of a PU ($\rm p$) transmitting its packets to a primary destination ($\rm d_{p}$). In addition, there exists an SU (${\rm s}$) communicating with a secondary destination ($\rm d_{s}$). The PU is plugged to a reliable power supply. This means that it never lacks energy/power. However, the SU is equipped with a rechargeable battery. We assume that the SU harvests energy from the surrounding environmental energy sources, e.g., solar energy, wind energy, etc. All nodes are equipped with infinite capacity buffers to store fixed-length ($\beta$ bits) data packets. Time is slotted and the duration of one time slot is $T$ seconds. The transmission of a packet takes one time slot.

The channel between every transmitter-receiver pair exhibits frequency-flat Rayleigh block fading, i.e., the channel coefficient remains constant for one time slot and changes independently from a slot to another. Moreover, a zero-mean additive white Gaussian noise of power spectral density $\mathcal{N}_{\circ}$ Watts/Hz is taken into account. The channel coefficient between transmitter $\rm i$ and receiver $\rm j$ at the $n$th time slot is denoted by $h_{\rm i,j}[n]$. According to the Rayleigh fading assumption, $h_{\rm  i,j}[n]$ is a complex Gaussian zero-mean random variable with variance $ \sigma_{\rm i,j}^{2}$, i.e., $h_{\rm i,j}[n] \sim \mathcal{CN}(0,\sigma_{\rm i,j}^{2})$. Therefore, $|h_{\rm i,j}[n]|^2$ is exponentially-distributed with rate $1/\sigma_{\rm i,j}^{2}$, i.e., $|h_{\rm i,j}[n]|^2 \sim \text{exp}(1/\sigma_{\rm i,j}^{2})$. All links are considered statistically independent.

Instead of assuming a simple packet-erasure model, we consider simultaneous multi-packet transmission capability. This opens room for both the PU and SU to use the spectrum concurrently. Assuming that the receivers are equipped with multi-packet reception (MPR) capability, transmitted data packets can survive the interference caused by concurrent transmissions if the received signal-to-interference-and-noise ratio (SINR) exceeds the threshold required for successful decoding. The SU performs probabilistic spectrum sensing. We take into account erroneous spectrum sensing outcomes.

Next, we present the queueing model of the system followed by the description of the proposed spectrum access protocol.

\subsection{Queueing Model}
\begin{figure}[t]
\begin{center}
\includegraphics[width=1\columnwidth , height=0.67\columnwidth]{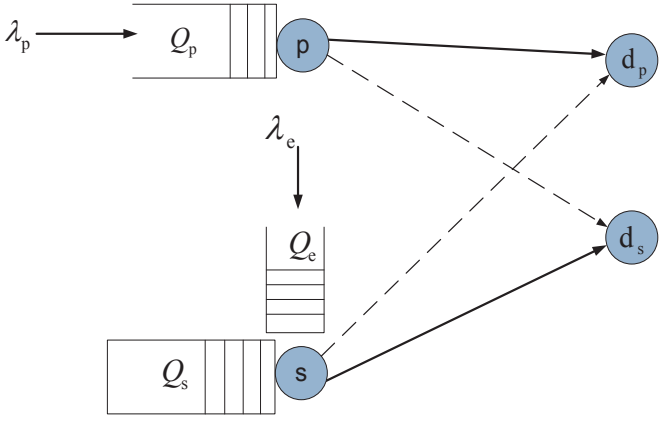}
\caption{Cognitive radio network under consideration.} \label{Fig1}
\end{center}
\vspace{-5mm}
\end{figure}

The queues involved in system analysis, shown in Fig. \ref{Fig1}, are described as follows:
\begin{itemize}
\item $Q_{\rm p}$: stores the data packets of the PU.

\item $Q_{\rm s}$: stores the data packets of the SU.

\item $Q_{\rm e}$: stores the harvested energy at the SU.
\end{itemize}

The bursty nature of information sources is taken into account through modelling the data arrivals at the PU as a Bernoulli process with rate $\lambda_{\rm p}$ (packets/slot). In other words, at any given time slot, a packet arrives at the PU with probability $\lambda_{\rm p}<1$. On the other hand, we assume that $Q_{\rm s}$ is backlogged, i.e., the SU always has packets awaiting transmission to $\rm d_s$. The arrivals  at the energy queue are assumed to follow a stationary Poisson process with rate $\lambda_{\rm e}$. This captures the random availability of ambient energy sources. The arrival processes at $Q_{\rm p}$ and $Q_{\rm e}$ are independent of each other, and are independent and identically-distributed (i.i.d) across time slots. Upon successful reception of a primary packet, $\rm  d_{p}$ broadcasts an acknowledgement (ACK). However, if $\rm d_{p}$ fails to decode a primary packet, it broadcasts a negative-acknowledgement (NACK). ACKs and NACKs are assumed instantaneous and can be heard by the PU and SU. The PU receives the feedback sent by $\rm d_{p}$ reliably with probability $1$. On the contrary, the SU overhears reliable primary feedback with probability $q$.

The instantaneous evolution of queue lengths is captured as
\begin{align}\label{queue evolution}
Q_{\rm i}[n+1]= \left(Q_{\rm i}[n] - L_{\rm i}[n]\right)^{+} + \mathcal{A}_{\rm i}[n], ~\rm i \in \{\rm p,e\}
\end{align}
where $(z)^{+}=\text{max}(z,0)$ and $Q_{\rm i}[n]$ denotes the number of packets in the $i$th queue at the beginning of the $n$th time slot. $L_{\rm i}[n]$ and $\mathcal{A}_{\rm i}[n]$ denote the departures and arrivals corresponding to the $\rm i$th queue in the $n$th time slot, respectively.

\subsection{Spectrum Access Protocol}\label{system_model_B}
Whenever $Q_{\rm p}$ is non-empty, the PU transmits a packet with average power $P_{\rm p}$. However, the SU probably attempts to transmit the head-of-line packet in its data queue if its battery has got the energy required to support that transmission. The spectrum access protocol employed at the SU exploits the available primary feedback signals. The SU may overhear nothing if there is no primary transmission. However, it may overhear an ACK or a NACK if the primary receiver correctly decodes, or fails to decode, the primary transmission, respectively, while the SU decodes the feedback signals correctly. Otherwise, the SU may overhear un-decodable feedback signals. Given that a primary transmission takes place in a given time slot, the SU is capable of decoding the primary feedback signals with probability $q$ as indicated earlier. If the SU receives nothing, an ACK or un-decodable feedback, it gains no information about the PU's activity in the next time slot. Therefore, the SU operates as follows.
\begin{itemize}
\item If the energy required for transmission is available at $Q_{\rm e}$, it performs spectrum sensing with probability $\alpha_{\rm s}$ from the beginning of the time slot for a duration $\tau < T$ seconds to detect the possible activity of the PU.

\item The SU transmits with probability $\alpha_{\rm f}$ or $\alpha_{\rm b}$ if the PU is detected to be idle or busy, respectively.

\item At the beginning of the time slot, if the SU decides not to sense the spectrum (which happens with probability $1-\alpha_{\rm s}$), it immediately decides whether to transmit with probability $\alpha_{\rm t}$ or to remain idle for the rest of the time slot with probability $1-\alpha_{\rm t}$.
\end{itemize}
Whenever the SU transmits in these cases, it uses an average power $P_{\rm s}^{\left(1\right)}$ when the PU is sensed to be inactive; and $P_{\rm s}^{\left(2\right)}$ when the PU is sensed to be active. Note that $P_{\rm s}^{\left(2\right)}$ should be lower than $P_{\rm s}^{\left(1\right)}$ to reduce the expected interference on the primary transmission. On the other hand, if a NACK is overheard by the SU, it knows that the PU will retransmit the lost packet during the next time slot with probability $1$. Being sure that the PU is active in the next slot, the SU does not need to perform spectrum sensing. Therefore, it accesses the channel from the beginning of the time slot with probability $\alpha_{\rm r}$. In that case, the SU transmits with average power $P_{\rm s}^{\left(3\right)} ~\le~ P_{\rm s}^{\left(2\right)}~\le~  P_{\rm s}^{\left(1\right)}$ to reduce its interference to the primary receiver.

We assume that energy dissipation from $Q_{\rm e}$ happens when the SU transmits data. The energy consumed in spectrum sensing and primary feedback decoding is negligible and out of the scope of this paper.

\begin{figure}[t]
\begin{center}
\includegraphics[width=1\columnwidth , height=0.67\columnwidth]{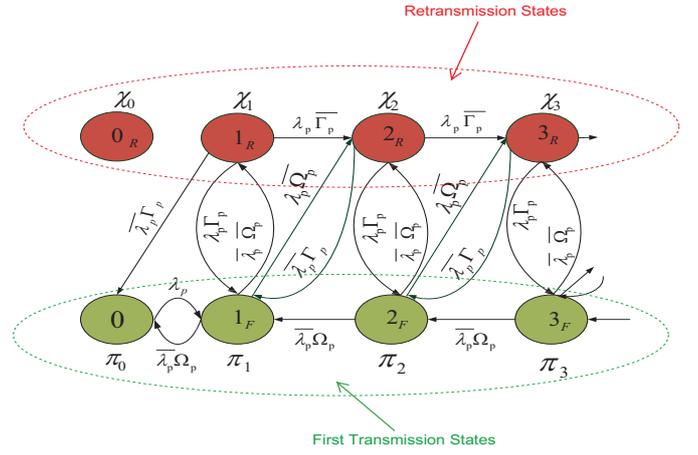}
\caption{Markov chain of the PU's queue-length evolution.} \label{Fig2}
\end{center}
\vspace{-5mm}
\end{figure}

\section{Performance Analysis}\label{performance_analysis}
In this section, we provide the queueing analysis of the proposed access scheme. In particular, we derive stability conditions on the queues involved in the system. The system is considered stable when all of its queues are stable. Queue stability is loosely defined as having a bounded queue size, i.e., the number of packets in the queue does not grow to infinity as time progresses \cite{sadek}. Furthermore, we provide a closed-form expression for the average primary packet delay.

\subsection{Queueing Analysis}

The PU's queue-length evolution Markov chain (MC) is shown in Fig. \ref{Fig2}\footnote{State-self transitions are not depicted on the graph for visual clarity. Throughout the paper,
$\overline{y}=1-y$.}. We refer to the probability of the queue having $\rm k$ packets and transmitting for the first time as $\pi_{\rm k}$, where $F$ in Fig. \ref{Fig2} denotes the first transmission. The probability of the queue having $\rm k$ packets and retransmitting is $\chi_{\rm k}$, where $R$ in Fig. \ref{Fig2} denotes retransmission. The probability of successful transmission of a PU's packet in case of first transmission and a retransmission is given by $\Omega_{\rm p}$ and $\Gamma_{\rm p}$, respectively. We proceed with calculating these probabilities.
\begin{equation}\label{alpha_p}
\Omega_{\rm p}=\mathbb{P} \gamma+\overline{\mathbb{P}}\mathbf{P}_{\rm 0}(P_{\rm p},0)
\end{equation}
where $\mathbb{P}$ is the probability that $Q_{\rm e}$ has got the amount of energy sufficient for one secondary packet transmission, $\mathbf{P}_{\rm i}(P_{\rm A},P_{\rm B})$ is derived in the Appendix and $\gamma$ is given by
\begin{align}\label{gamma}
\gamma  &= \overline{\alpha_{\rm s}}\bigg\{\alpha_{\rm t}\mathbf{P}_{\rm 0}(P_{\rm p},P_{\rm s}^{\left(1\right)}) + \overline{\alpha_{\rm t}}\mathbf{P}_{\rm 0}(P_{\rm p},0)\bigg\}\notag \\
&+\alpha_{\rm s}{\rm P}_{\rm MD}\bigg\{\alpha_{\rm f}\mathbf{P}_{\rm 0}(P_{\rm p},P_{\rm s}^{\left(1\right)})+ \overline{\alpha_{\rm f}}\mathbf{P}_{\rm 0}(P_{\rm p},0)\bigg\} \notag \\
&+\alpha_{\rm s}\overline{{\rm P}_{\rm MD}}\bigg\{\alpha_{\rm b}\mathbf{P}_{\rm 0}(P_{\rm p},P_{\rm s}^{\left(2\right)}) + \overline{\alpha_{\rm b}}\mathbf{P}_{\rm 0}(P_{\rm p},0)\bigg\}.
\end{align}
We denote the probability of miss-detection by ${\rm P}_{\rm MD}$. On the other hand, it can be shown that $\Gamma_{\rm p}$ is given by
\begin{table}
\renewcommand{\arraystretch}{2}
\begin{center}
\begin{tabular}{ |c |c|  }
    \hline
    $\eta$ & $\lambda_{\rm p}\Omega_{\rm p}+\left(1-\lambda_{\rm p}\right)\Gamma_{\rm p}$ \\[5pt]\hline
  $\pi_{\rm \circ}$ & $\frac{\eta-\lambda_{\rm p}}{\Gamma_{\rm p}}$ \\[5pt]\hline
  $\chi_{\rm \circ}$ & 0 \\[5pt]\hline
  $\pi_1$& $\pi_{\rm \circ}\tfrac{\lambda_{\rm p}}{1-\lambda_{\rm p}}\frac{\lambda_{\rm p}+\left(1-\lambda_{\rm p}\right)\Gamma_{\rm p}}{\eta}$  \\[5pt]\hline
  $\chi_1$ & $\pi_{\rm \circ}\frac{\lambda_{\rm p}}{\eta}\left(1-\Omega_{\rm p}\right)$ \\ [5pt]\hline
  $\pi_{\rm k},{\rm k}\geq 2$ & $\pi_{\rm \circ}\frac{\lambda_{\rm p}\left(1-\Omega_{\rm p}\right)}{\left(1-\eta\right)^2} \bigg[ \frac{\lambda_{\rm p} \left(1-\eta\right)}{\left(1-\lambda_{\rm p}\right) \eta}  \bigg]^{\rm k}$ \\[5pt]\hline
  $\chi_{\rm k},{\rm k}\geq 2$ & $\pi_{\rm \circ}\frac{\left(1-\lambda_{\rm p}\right)\left(1-\Omega_{\rm p}\right)}{\left(1-\eta\right)^2} \bigg[ \frac{\lambda_{\rm p} \left(1-\eta\right)}{\left(1-\lambda_{\rm p}\right) \eta}  \bigg]^{\rm k}$ \\[5pt]\hline
  $\sum_{\rm k=1}^{\infty}\pi_{\rm k}$ & $\pi_{\rm \circ}\frac{\lambda_{\rm p}\Gamma_{\rm p}}{\eta-\lambda_{\rm p}}=\lambda_{\rm p}$\\[5pt]\hline
   $\sum_{\rm k=1}^{\infty}\chi_{\rm k}$ & $\pi_{\rm \circ}\frac{\lambda_{\rm p}}{\eta-\lambda_{\rm p}}\left(1-\Omega_{\rm p}\right)=\frac{\lambda_{\rm p} }{\Gamma_{\rm p}}\left(1-\Omega_{\rm p}\right)$
  \\[5pt]\hline
\end{tabular}
\vspace{+4mm}
\caption{Steady-state distribution for the PU's MC.}
\label{table}
\end{center}
\vspace{-5mm}
\end{table}
\begin{equation}\label{Gamma_p}
\Gamma_{\rm \!p}\!=\!\mathbb{P}\bigg[q\bigg\{\!\alpha_{\rm r}\mathbf{P}_{\rm 0}(P_{\rm p},P_{\rm s}^{\left(3\right)})
\!+\!\overline{\alpha_{\rm r}}\mathbf{P}_{\rm 0}(P_{\rm p},0)\!\bigg\}\!+\!\overline{q}\gamma
\bigg]\!+\!\overline{\mathbb{P}}\mathbf{P}_{\rm 0}(P_{\rm p},0).
\end{equation}
We solve for the steady-state distribution of the PU's MC.
Solving the state balance equations of the MC depicted in Fig. \ref{Fig2}, we obtain the state probabilities which are provided in Table \ref{table}. The probability $\pi_0$ is obtained using the normalization condition $\sum_{\rm k=0}^{\infty} (\pi_{\rm k}+\chi_{\rm k})=1$.
It should be noticed that $\lambda_{\rm p} < \eta$, where $\eta$ is defined in Table \ref{table}, is a condition for the sum $\sum_{\rm k=0}^{\infty} (\pi_{\rm k}+\chi_{\rm k})$ to exist. This condition ensures the existence of a stationary distribution for the MC. Furthermore, it guarantees the stability of $Q_{\rm p}$, i.e., $Q_{\rm p}$ has a non-zero probability of being empty ($\pi_{\rm 0}>0$).

From (\ref{alpha_p}) and (\ref{Gamma_p}), we notice that $\Omega_{\rm p}$ and $\Gamma_{\rm p}$ depend on the state of $Q_{\rm e}$ through the term $\mathbb{P}$. Therefore, we need to model the energy dissipation from the battery of the SU, i.e., $Q_{\rm e}$. Let ${\rm P}_{\rm FA}$ denote the probability of false alarm. According to the proposed access protocol described in Section \ref{system_model_B}, the mean service rate of $Q_{\rm e}$ is given in (\ref{mu_e}) at the top of the following page.

                \begin{figure*}[!t]
\normalsize
\setcounter{equation}{4}
\begin{align}\label{mu_e}
\mu_{\rm e}\!&=\!\pi_{\rm 0}\bigg\{\overline{\alpha_{\rm s}} \alpha_{\rm t} P_{\rm s}^{\left(1\right)} T\! +\! \bigg[\alpha_{\rm s}\alpha_{\rm f}\overline{{\rm P}_{\rm FA}}
P_{\rm s}^{\left(1\right)}+\alpha_{\rm s}\alpha_{\rm b}{\rm P}_{\rm FA} P_{\rm s}^{\left(2\right)}\bigg]T_{\rm s}\bigg\}\!+\!
\left(\displaystyle\sum_{\rm k=1}^{\infty}\!\pi_{\rm k}\right)\!\!\bigg\{ \overline{\alpha_{\rm s}}\alpha_{\rm t}P_{\rm s}^{\left(1\right)}T
\!+\!
\alpha_{\rm s}\bigg[\alpha_{\rm f}{\rm P}_{\rm MD} P_{\rm s}^{\left(1\right)}\!+\!\alpha_{\rm b}\overline{{\rm P}_{\rm MD}}P_{\rm s}^{\left(2\right)}\bigg]T_{\rm s} \bigg\} \notag \\ &\,\,\,\,\,\,\,\,\
 +\left(\!\displaystyle\sum_{\rm k=1}^{\infty}\!\chi_{\rm k}\!\!\right)\!
\bigg\{\!\!q\alpha_{\rm r}P_{\rm s}^{\left(3\right)}T\!\!+\!\!\overline{q} \!\bigg(\! \overline{\alpha_{\rm s}}\alpha_{\rm t} P_{\rm s}^{\left(1\right)} T \!+\!
\alpha_{\rm s}\bigg[\!\alpha_{\rm f}{\rm P}_{\rm MD} P_{\rm s}^{\left(1\right)}\!+\!\alpha_{\rm b}\overline{{\rm P}_{\rm MD}}P_{\rm s}^{\left(2\right)}\!\bigg]\!T_{\rm s}\!\!\bigg)\!\!\bigg\}, \ \text{where} \ T_{\rm s}=T-\tau.
\\
\label{mu_s}
\mu_{\rm s}\! &=\!\mathbb{P}\bigg[ \pi_{\rm 0}\bigg\{\overline{\alpha_{\rm s}} \alpha_{\rm t} \mathbf{P}_{\rm \!0}(P_{\rm s}^{\left(1\right)},0)\! +\! \alpha_{\rm s}\big[\alpha_{\rm f}\overline{{\rm P}_{\rm FA}}
\mathbf{P}_{\rm \!1}(P_{\rm s}^{\left(1\right)},0)+\alpha_{\rm b}{\rm P}_{\rm FA}\mathbf{P}_{\rm \!1}(P_{\rm s}^{\left(3\right)},0) \big]\!\bigg\} \notag \\
&\,\,\,\,\,\,\,\,\  +\!
\left(\!\displaystyle\sum_{\rm k=1}^{\infty}\!\!\pi_{\rm k}\!\right)\!\!\!\bigg\{\! \overline{\alpha_{\rm s}}\alpha_{\rm t}
\mathbf{P}_{\rm \!0}\!(P_{\rm s}^{\left(1\right)},P_{\rm p})
\!\!+\!\!
\alpha_{\rm s}\big[\alpha_{\rm f}{\rm P}_{\rm MD} \mathbf{P}_{\rm \!1}\!(P_{\rm s}^{\left(1\right)},P_{\rm p})+\alpha_{\rm b}\overline{{\rm P}_{\rm MD}}\mathbf{P}_{\rm \!1}\!(P_{\rm s}^{\left(3\right)},P_{\rm p})\big]\! \bigg\}
\notag \\ &\,\,\,\,\,\,\,\,\
+\!\left(\displaystyle\sum_{\rm k=1}^{\infty}\chi_{\rm k}\right)
\bigg\{ q \alpha_{\rm r}\mathbf{P}_{\rm 0}(P_{\rm s}^{\left(2\right)},P_{\rm p})
+\overline{q}\bigg(\overline{\alpha_{\rm s}}\alpha_{\rm t}\mathbf{P}_{\rm 0}(P_{\rm s}^{\left(1\right)},P_{\rm p})+
\alpha_{\rm s}\big[\alpha_{\rm f}{\rm P}_{\rm MD} \mathbf{P}_{\rm 1}(P_{\rm s}^{\left(1\right)},P_{\rm p})+\alpha_{\rm b}\overline{{\rm P}_{\rm MD}} \mathbf{P}_{\rm 1}(P_{\rm s}^{\left(3\right)},P_{\rm p})\big]
\!\bigg)\!\bigg\}\bigg]\!.
\end{align}
\hrulefill
\end{figure*}
It is obvious that the service rate of $Q_{\rm e}$ depends on the state of $Q_{\rm p}$, i.e., $\pi_{\rm k}$ and $\chi_{\rm k}$. Thus, $Q_{\rm p}$ and $Q_{\rm e}$ are two interacting queues. The relaxation of this interaction and the computation of $\mathbb{P}$ is provided in the next section.

We proceed next with characterizing the SU throughput, i.e., the mean service rate of $Q_{\rm s}$. It is given in (\ref{mu_s}) at the top of the following page.
The dependence of the SU throughput on both $Q_{\rm e}$ and $Q_{\rm p}$ is highlighted in (\ref{mu_s}). The SU transmits only when the energy required to support its transmission is available in $Q_{\rm e}$. This explains the role of $\mathbb{P}$ in (\ref{mu_s}). In addition, the SU's behavior depends on the state of the PU, i.e., the PU being in a first transmission or a retransmission state. By behavior here we refer to the decisions made by the SU concerning spectrum sensing and access and the choice of transmission powers.

\subsection{Average Primary Packet Delay}
Applying Little's law \cite{Bertsekas}, we obtain a closed-form expression for the average primary packet delay, which is given by
\begin{equation}
D_{\rm p}=\frac{1}{\lambda_{\rm p}}\sum_{\rm k=1}^{\infty} {\rm k} \left(\pi_{\rm k}+\chi_{\rm k}\right).
\end{equation}
\noindent Using the state probabilities provided in Table \ref{table},
\begin{equation}\label{delay_p}
\begin{split}
D_{\rm p} &=\frac{(\Omega_{\rm p}-\eta)(\eta-\lambda_{\rm p})^2+\left(1-\lambda_{\rm p}\right)^2 \left(1-\Omega_{\rm p}\right) \eta}{(\eta-\lambda_{\rm p})\left(1-\lambda_{\rm p}\right) \left(1-\eta\right) \Gamma_{\rm p}}.
  \end{split}
\end{equation}

\section{Bounds on Secondary Throughput and Problem Formulation}\label{lower_and_upper_bounds}
In this section, we derive lower and upper bounds on the throughput of the SU.
Our main goal now is to relax the interaction between $Q_{\rm e}$ and $Q_{\rm p}$ to be able to compute $\mathbb{P}$ and $\{\pi_{\rm k},\chi_{\rm k}\}_{\rm k=0}^{\infty}$ and hence, the throughput of the SU can be characterized.
\subsection{Lower Bound on $\mu_{\rm s}$}\label{lb_on_mu_s}
We analyze the effect of the states of $Q_{\rm p}$ and $Q_{\rm e}$ on the SU throughput. We begin first with $Q_{\rm p}$. If the PU's service rate is decreased, less number of packets probably depart $Q_{\rm p}$. This implies that the probability of $Q_{\rm p}$ being empty is lowered. Therefore, the PU is more likely to be active and hence, the interference on the SU is increased which lowers its throughput. This motivates us to derive lower bounds on $\Omega_{\rm p}$ and $\Gamma_{\rm p}$ provided in (\ref{alpha_p}) and (\ref{Gamma_p}), respectively. We note that the worst case scenario with respect to the PU occurs when it experiences continuous possible interference from the SU. Since the SU is assumed backlogged, continuous SU transmission possibly occurs when it always has the required energy, i.e., $\mathbb{P}=1$. Substituting by $\mathbb{P}=1$ in (\ref{alpha_p}) and (\ref{Gamma_p}),
\begin{eqnarray}
\Omega_{\rm p} &\geq & \gamma \label{alpha_p_lb} \\
\Gamma_{\rm p} &\geq & q\bigg\{\alpha_{\rm r}\mathbf{P}_{\rm 0}(P_{\rm p},P_{\rm s}^{\left(3\right)})
+\overline{\alpha_{\rm r}}\mathbf{P}_{\rm 0}(P_{\rm p},0)\bigg\}+\overline{q}\gamma \label{Gamma_p_lb}
\end{eqnarray}
where $\gamma$ is given by (\ref{gamma}).

On the other hand, we note from (\ref{mu_e}) that the energy dissipation from $Q_{\rm e}$, whenever a secondary transmission occurs, is one of four levels: (i) $P_{\rm s}^{\left(1\right)} T$, (ii) $P_{\rm s}^{\left(1\right)}T_{\rm s}$, (ii) $P_{\rm s}^{\left(2\right)}T_{\rm s}$ and (iii) $P_{\rm s}^{\left(3\right)}T$. We assume that $Q_{\rm e}$ dissipates $P_{\rm s}^{\left(1\right)} T$ for each secondary transmission provided that this energy is available in $Q_{\rm e}$. This provides an upper bound on the service rate of $Q_{\rm e}$ because $P_{\rm s}^{\left(1\right)} T$ is the one which has the highest level among the aforementioned four levels. Hence, the SU battery is depleted faster which negatively impacts its throughput. This assumption renders $Q_{\rm e}$ an $\text{M}/\text{D}/1$ queue with arrival rate $\lambda_{\rm e}$ and service rate $P_{\rm s}^{\left(1\right)} T$. Therefore, $\mathbb{P}$ is \cite{krikidis2012stability}
\begin{equation}\label{P_Qe_ub}
\mathbb{P}=\text{min}\left[1,\frac{\lambda_{\rm e}}{P_{\rm s}^{\left(1\right)} T}\right].
\end{equation}

We use the lower bounds on $\Omega_{\rm p}$ and $\Gamma_{\rm p}$ provided on the right hand sides of (\ref{alpha_p_lb}) and (\ref{Gamma_p_lb}), respectively, to compute the state probabilities of the PU's MC given by Table \ref{table}. Substituting by these state probabilities along with $\mathbb{P}$ as given by (\ref{P_Qe_ub}) in (\ref{mu_s}), we get a lower bound on the throughput of the SU.

\subsection{Upper Bound on $\mu_{\rm s}$}\label{ub_on_mu_s}
We use a similar approach to the one presented in Section \ref{lb_on_mu_s} to derive an upper bound on the SU throughput. When the service rate of the PU's queue increases, this increases the availability of time slots in which the PU is silent. Thus, the interference on the SU decreases and its throughput is enhanced. This motivates us to derive upper bounds on $\Omega_{\rm p}$ and $\Gamma_{\rm p}$.
The best case scenario with respect to the PU occurs when it experiences no interference from the SU, as if the SU does not exist. This happens when the battery of the SU is always empty, corresponding to $\mathbb{P}=0$. Substituting by $\mathbb{P}=0$ in (\ref{alpha_p}) and (\ref{Gamma_p}), we get
\begin{eqnarray}
\Omega_{\rm p} & \leq & \mathbf{P}_{\rm 0}(P_{\rm p},0) \label{alpha_p_ub} \\
\Gamma_{\rm p} & \leq & \mathbf{P}_{\rm 0}(P_{\rm p},0) \label{Gamma_p_ub}.
\end{eqnarray}

On the other hand, we assume that $Q_{\rm e}$ dissipates $\text{min}[P_{\rm s}^{\left(2\right)}T_{\rm s},P_{\rm s}^{\left(3\right)}T]$ for each secondary transmission provided that this energy is available in $Q_{\rm e}$. This provides a lower bound on the service rate of $Q_{\rm e}$. Hence, the SU battery lifetime increases which positively impacts its throughput. This assumption renders $Q_{\rm e}$ an $\text{M}/\text{D}/1$ queue with arrival rate $\lambda_{\rm e}$ and service rate $\text{min}[P_{\rm s}^{\left(2\right)}T_{\rm s},P_{\rm s}^{\left(3\right)}T]$. Therefore, $\mathbb{P}$ is given by \cite{krikidis2012stability}
\begin{equation}\label{P_Qe_lb}
\mathbb{P}=\text{min}\left[1,\frac{\lambda_{\rm e}}{\text{min}[P_{\rm s}^{\left(2\right)}T_{\rm s},P_{\rm s}^{\left(3\right)}T]}\right].
\end{equation}

We use the upper bounds on $\Omega_{\rm p}$ and $\Gamma_{\rm p}$ provided on the right hand sides of (\ref{alpha_p_ub}) and (\ref{Gamma_p_ub}), respectively, to compute the state probabilities of the PU's MC given by Table \ref{table}. Substituting by these state probabilities along with $\mathbb{P}$ as given by (\ref{P_Qe_lb}) in (\ref{mu_s}), we get an upper bound on the SU throughput.

\subsection{Problem Formulation}
Using the relevant mean service rates under each of the proposed bounds, for a fixed $\lambda_{\rm p}$, the following constrained optimization problem is solved numerically
\begin{align}\label{optimization}
& \underset{\alpha_{\rm s},\alpha_{\rm f},\alpha_{\rm t},\alpha_{\rm b},\alpha_{\rm r},P_{\rm s}^{\left(1\right)},P_{\rm s}^{\left(2\right)},P_{\rm s}^{\left(3\right)}}{\text{maximize}}
& & \mu_{\rm s} \notag \\
& ~~~~~ \text{subject to}
& & 0 \leq \alpha_{\rm s},\alpha_{\rm f},\alpha_{\rm t},\alpha_{\rm b},\alpha_{\rm r} \leq 1 \notag \\
&&& 0 \leq P_{\rm s}^{\left(1\right)},P_{\rm s}^{\left(2\right)},P_{\rm s}^{\left(3\right)} \leq P_{\rm \circ} \notag \\
&&& \lambda_{\rm p} < \eta \notag \\
&&& D_{\rm p} \leq D_{\rm \circ}.
\end{align}
We set maximum secondary transmission power constraint $P_{\rm \circ}$, and a threshold $D_{\rm \circ}$ below which the average primary packet delay is kept. Note that the optimization problem is solved at the secondary terminal. We use MatLab's fmincon to solve the optimization problems as in \cite{wimob,6568963,6177245,4472701,6648968} and the references therein.

\section{Numerical results}\label{numerical_results}
In this section, the performance of the proposed scheme is evaluated in terms of the SU throughput. We formulate and solve an optimization problem with the objective of maximizing the secondary throughput subject to certain guaranteed QoS measures for the PU. We solve for the optimal sensing and access probabilities as well as the average secondary transmission powers that maximize the SU throughput while simultaneously keeping the PU's queue stable and the primary packet delay below a certain threshold.

We show our results for packet-length $\beta=10$ bits, time slot duration $T=1$ second and sensing duration $\tau=0.3$ second. Sensing errors are taken into account through setting ${\rm P}_{\rm MD}={\rm P}_{\rm FA}=0.3$. All links are considered statistically equivalent where $\sigma_{\rm i,j}^{2}=1$ $\forall {\rm i} \in \{\rm p,s\}$ and ${\rm j} \in \{\rm d_{p}, d_{s}\}$. The bandwidth of these channels is set to $W=8$ Hz. The PU transmits its packets with average power $P_{\rm p}=20$ Watts. The power spectral density of noise is normalized to unity, i.e., $\mathcal{N}_{\rm \circ}=1$ Watts/Hz. We choose $P_{\rm \circ}=32$ Watts (corresponding to a signal-to-noise-ratio $=6$ dB if the SU is transmitting alone).
As indicated earlier, $\lambda_{\rm p}<\eta$ is the constraint that guarantees the stability of the PU's queue.

\begin{figure}[t]
\begin{center}
\includegraphics[width=1\columnwidth]{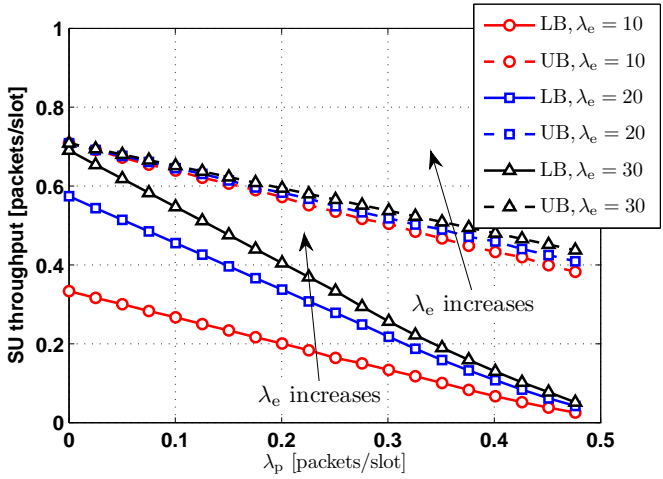}
\caption{Optimal lower and upper bounds on $\mu_{\rm s}$ versus $\lambda_{\rm p}$ for different values of $\lambda_{\rm e}$.} \label{Fig3}
\end{center}
\vspace{-5mm}
\end{figure}

\begin{figure}[t]
\begin{center}
\includegraphics[width=1\columnwidth]{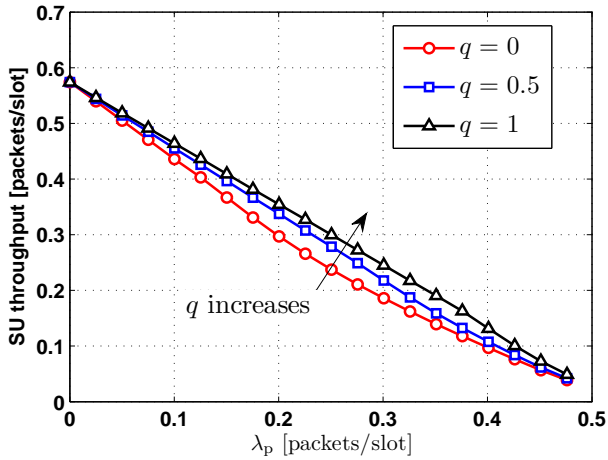}
\caption{Optimal lower bound on $\mu_{\rm s}$ versus $\lambda_{\rm p}$ for different values of $q$.} \label{Fig4}
\end{center}
\vspace{-5mm}
\end{figure}
\begin{figure}[t]
\begin{center}
\includegraphics[width=1\columnwidth]{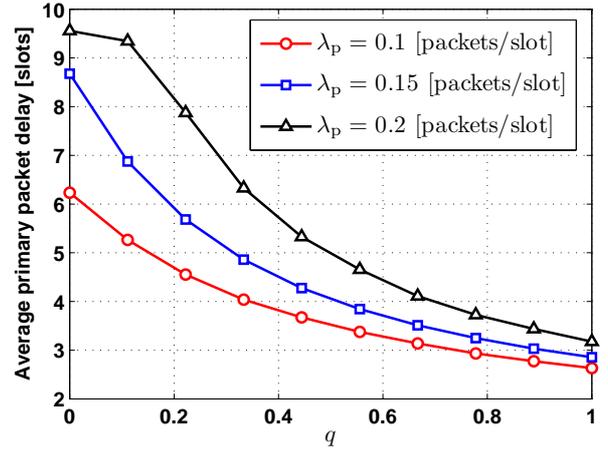}
\caption{Optimal upper bound on $D_{\rm p}$ versus $q$ for different values of $\lambda_{\rm p}$.} \label{Fig5}
\end{center}
\vspace{-5mm}
\end{figure}

In Fig. \ref{Fig3}, we plot the resulting solution of (\ref{optimization}), with $\mu_{\rm s}$ given by the lower and upper bounds explained in Sections \ref{lb_on_mu_s} and \ref{ub_on_mu_s}, respectively, versus $\lambda_{\rm p}$ at a chosen value of $q=0.5$ and $D_{\rm \circ}=10$ time slots. This corresponds to a PU running a delay-tolerant application. In the figure's legend, we denote the upper bound and lower bound by `UB' and `LB', respectively. We investigate the effect of the energy arrival rate at $Q_{\rm e}$ on the SU throughput through plotting the results at different values of $\lambda_{\rm e}$. As we expected, the figure shows that the SU throughput is enhanced as $\lambda_{\rm e}$ increases. This is attributed to the fact that the restriction on the SU's activity originating from its energy resources is relaxed as $\lambda_{\rm e}$ increases. From the figure, we note that the gap between the inner and the outer bound decreases with the increase of $\lambda_{\rm e}$.

Next, we depict the effect of unreliable primary feedback signals on SU throughput and average primary packet delay. Towards this objective, we plot the optimal lower bound on $\mu_{\rm s}$ versus $\lambda_{\rm p}$ for different values of $q$ in Fig. \ref{Fig4}. We choose $\lambda_{\rm e}=20$ and $D_{\rm \circ}=10$. As shown in the figure, the throughput of the SU is enhanced as $q$ increases. This shows that the best case scenario occurs when the SU receives perfect primary feedback, i.e., $q=1$. However, the worst case scenario occurs when the SU overhears no reliable feedback from the PU, i.e., $q=0$. The case of $q=0$ boils down to the case in which the SU can not overhear/decode the primary feedback signals.
On the other hand, we evaluate the underlying average PU delay.
We use the resulting optimal sensing and access probabilities as well as average secondary transmission powers, used to plot Fig. \ref{Fig4}, in computing the average primary packet delay given by (\ref{delay_p}). It is worth noting that the primary delay computed through solving (\ref{optimization}) with the objective of maximizing the lower bound on $\mu_{\rm s}$ is an upper bound on the achievable PU delay. In Fig. \ref{Fig5}, we plot the PU's delay versus $q$ at selected values of $\lambda_{\rm p}$. Thus, it can be seen that the primary delay is a decreasing function of $q$. Therefore, we reach the conclusion that reliable feedback signals overheard by the SU is not only in its interest, but it also enhances the PU's performance.

\begin{figure}[t]
\begin{center}
\includegraphics[width=1\columnwidth]{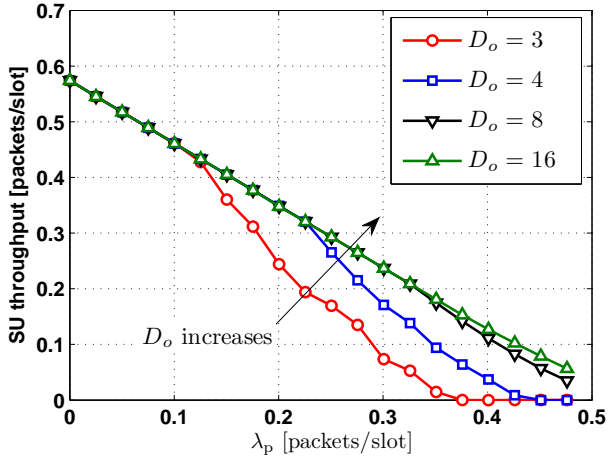}
\caption{Optimal lower bound on $\mu_{\rm s}$ versus $\lambda_{\rm p}$ for different values of delay constraint.} \label{Fig6}
\end{center}
\vspace{-5mm}
\end{figure}

Furthermore, we investigate the effect of the QoS measures guaranteed for the PU on the throughput of the SU. Specifically, we depict the effect of varying the average primary queueing delay constraint on the SU throughput in Fig. \ref{Fig6}. The results are plotted for $q=0.8$ and $\lambda_{\rm e}=20$ energy packets/slot. From Fig. \ref{Fig5}, we note that the lower bound on SU throughput is enhanced as the primary delay constraint is relaxed, i.e., $D_{\rm \circ}$ is increased. This intuitive result is attributed to the fact that for a less restrictive primary delay constraint, the feasible set of the optimization variables is widened. Thus, we are able to achieve better SU throughput at higher values of $D_{\rm \circ}$.

\begin{figure}[t]
\begin{center}
\includegraphics[width=1\columnwidth]{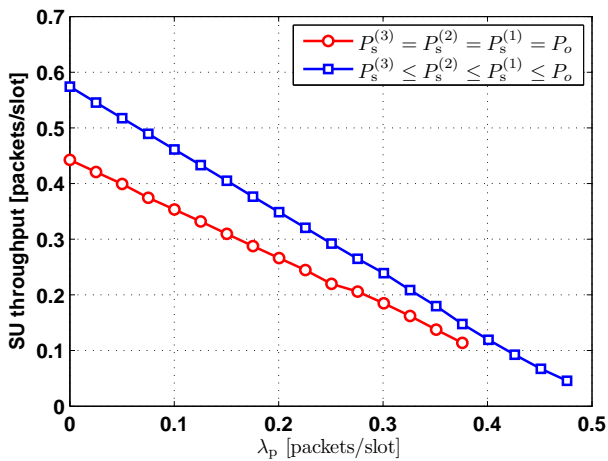}
\caption{Effect of power allocation on SU throughput.} \label{Fig7}
\end{center}
\vspace{-5mm}
\end{figure}

In Fig. \ref{Fig7}, we show the essence of performing power allocation at the SU. We plot the resulting optimal lower bound on $\mu_{\rm s}$ with and without optimizing over the powers $P_{\rm s}^{\left(1\right)}$, $P_{\rm s}^{\left(2\right)}$ and $P_{\rm s}^{\left(3\right)}$. In the latter case, we set $P_{\rm s}^{\left(1\right)}$, $P_{\rm s}^{\left(2\right)}$ and $P_{\rm s}^{\left(3\right)}$ to their maximum possible value, i.e., $P_{\rm \circ}$. We set $\lambda_{\rm e}=20$ energy packets/slot and $q=0.8$.
From the figure, we note that the SU throughput is enhanced when we optimize over the transmission powers and the resulting optimal values for $P_{\rm s}^{\left(1\right)}$, $P_{\rm s}^{\left(2\right)}$ and $P_{\rm s}^{\left(3\right)}$ is definitely less than $P_{\rm \circ}$. Thus, we achieve higher SU throughput at lower average power consumption. We note that at $\lambda_{\rm p}=0.3759$ packets/slot, the problem becomes infeasible and there are no optimization variables would satisfy the primary constraints. Hence, the SU remains silent and it gains no access to the channel.

\section{Conclusion}\label{conclusion}
We propose a novel random spectrum sensing and access scheme for a cognitive radio SU equipped with energy harvesting capability. The cognitive user utilizes the primary feedback signals to decide on its spectrum sensing and access probabilities. Furthermore, we optimize over the SU's transmission power with the objective of maximizing its throughput while simultaneously keeping the primary queue stable, and its average delay below a certain threshold. We investigate the effect of erroneous primary feedback signals overheard by the SU. We show that reliable feedback is in the interest of both the PU and SU. Moreover, we show the essence of performing power allocation at the SU compared to schemes in which SUs transmit with their maximum permissible power.

\section*{Appendix}
We derive the probability of successful transmission in the presence of interference. Consider two nodes, $\rm A$ and $\rm B$, transmitting to a common destination, $\rm d$, with average powers $P_{\rm A}$ and $P_{\rm B}$, respectively. With respect to $\rm d$, node's $\rm A$ transmission is the intended signal, while node $\rm B$ is considered an interferer. In a given time slot, $A$ transmits one packet ($\beta$ bits). The transmission duration is either $T-\tau$ or $T$ (seconds) depending on whether the transmission is preceded by a sensing phase or not, respectively. Thus, the transmission rate is given by
\begin{equation}
r_{\rm i}=\frac{\beta}{T-{\rm i}\tau}, ~{\rm i} \in \{0,1\}.
\end{equation}
Consequently, we use the notion of channel outage to write the probability of success on the link $\rm A \rightarrow d$ as
\begin{equation}
\mathbf{P}_{\rm \!i}(P_{\rm \!A},P_{\rm \!B})\!=\!\text{Pr}\! \left\{W\log\! \left[1\!+\!\frac{P_{\rm A}|h_{\rm A,d}|^2}{\mathcal{N}_{\circ}W+P_{\rm B}
|h_{\rm B,d}|^2}\right]\!>\!r_{\rm i}\right\}
\end{equation}
where $W$ is the channel bandwidth and $\text{Pr}\{\mathcal{E}\}$ denotes the probability of event $\mathcal{E}$. After some algebraic manipulation,
\begin{equation}
\mathbf{P}_{\rm \!i}(P_{\rm \!A},P_{\rm \!B})\!\!=\!\! \text{Pr}\! \left\{\! |h_{\rm A,d}|^2\!>\!\frac{2^{\frac{r_{\rm i}}{W}}-1}{P_{\rm \!A}/(\mathcal{N}_{\circ}W)}\!+\!\frac{(2^{\frac{r_{\rm i}}{W}}-1)P_{\rm \!B}|h_{\rm B,d}|^2}{P_{\rm \!A}}\! \right\}\!.
\end{equation}
For the ease of exposition, let $a=\frac{2^{\frac{r_{\rm i}}{W}}-1}{P_{\rm \!A}/(\mathcal{N}_{\circ}W)}$, $b=\frac{(2^{\frac{r_{\rm i}}{W}}-1)P_{\rm \!B}}{P_{\rm \!A}}$ and $X=|h_{\rm B,d}|^2$. Thus, we have
\begin{equation}\label{prob_success}
\mathbf{P}_{\rm \!i}(P_{\rm \!A},P_{\rm \!B})=\text{Pr} \left\{ |h_{\rm A,d}|^2>a+bX \right\}.
\end{equation}
Using total probability theory, (\ref{prob_success}) can be written as
\begin{equation}\label{total_prob}
\mathbf{P}_{\rm \!i}(P_{\rm \!A},P_{\rm \!B})=\int_{\rm 0}^{\infty} \text{Pr} \left\{ |h_{\rm A,d}|^2>a+bx \right\}
f_{\rm X}(x)dx
\end{equation}
where $f_{\rm X}(.)$ denotes the probability density function (PDF) of $X$. Following the channel model described in Section \ref{system_model}, $|h_{\rm A,d}|^2 \sim \text{exp}(1/\sigma_{\rm A,d}^{2})$ and $X \sim \text{exp}(1/\sigma_{\rm B,d}^{2})$. Therefore,
\begin{equation}\label{term_1}
\text{Pr} \left\{ |h_{\rm A,d}|^2>a+bx \right\}=e^{-(a+bx)/\sigma_{\rm A,d}^{2}}
\end{equation}
\begin{equation}\label{term_2}
f_{\rm X}(x)=\frac{1}{\sigma_{\rm B,d}^{2}}e^{-x/\sigma_{\rm B,d}^{2}}.
\end{equation}
Substituting by (\ref{term_1}) and (\ref{term_2}) in (\ref{total_prob}) and solving the integral, the probability of success on the link $A \rightarrow d$ in the presence of an interferer ($B$) is given by
\begin{equation}
\mathbf{P}_{\rm \!i}(P_{\rm \!A},P_{\rm \!B})=\frac{\sigma_{\rm A,d}^{2}~e^{-a/\sigma_{\rm A,d}^{2}}}{\sigma_{\rm A,d}^{2}+b~\sigma_{\rm B,d}^{2}}.
\end{equation}

\bibliographystyle
{IEEEtran}
\bibliography{IEEEabrv,energy_bib}
\balance
\end{document}